\documentclass[11pt]{article}
\usepackage{graphicx}
\usepackage{amssymb}
\def\lsim{\mathrel{\raise.3ex\hbox{$<$\kern-.75em\lower1ex\hbox{$\sim$}}}}
\def\gsim{\mathrel{\raise.3ex\hbox{$>$\kern-.75em\lower1ex\hbox{$\sim$}}}}

\newcommand{\be}{\begin{equation}}
\newcommand{\ee}{\end{equation}}
\newcommand{\bea}{\begin{eqnarray}}
\newcommand{\eea}{\end{eqnarray}}

\begin{document}

\thispagestyle{empty}

\title
{Challenging Lorentz noninvariant neutrino oscillations\\
without neutrino masses}
\author{V. Barger$^1$, D. Marfatia$^{2}$ and K. Whisnant$^3$\\[2ex]
\small\it $^1$Department of Physics, University of Wisconsin, Madison, WI 53706\\
\small\it $^2$Department of Physics and Astronomy, University of Kansas, Lawrence, KS 66045\\
\small\it $^3$Department of Physics, Iowa State University, Ames, IA 50011}

\date{}

\maketitle

\begin{abstract}

We show that the combined data from solar, long-baseline
and reactor neutrino experiments can exclude the generalized bicycle
model of Lorentz noninvariant direction-dependent and/or
direction-independent oscillations of massless neutrinos. This model
has five parameters, which is more than is needed in standard
oscillation phenomenology with neutrino masses. Solar data alone are
sufficient to exclude the pure direction-dependent case. The
combination of solar and long-baseline data rules out the
pure direction-independent case. With the addition of KamLAND data, a
mixture of direction-dependent and direction-independent terms in the
effective Hamiltonian is also excluded.

\end{abstract}

\newpage

\section{Introduction}

Neutrino oscillations are now a well-established
phenomenon~\cite{review}.  Data from solar, atmospheric, reactor and
accelerator experiments may be explained by the now standard scenario
with three active, massive neutrinos, with the possible exception of
the LSND experiment~\cite{LSND}. Recently it has been suggested that
Lorentz-invariance and $CPT$ violating interactions originating at the
Planck scale can also lead to neutrino oscillations,
with or without neutrino mass~\cite{Coleman,BPWW,K1,K2,Kat}.  
These interactions can be nonisotropic, which could lead to different
oscillation parameters for neutrinos propagating in different
directions.

The effective hamiltonian that describes the evolution of massless
neutrinos in vacuum in the presence of Lorentz-invariance violating
interactions may be written as~\cite{K1}
\be
(h_{eff})_{ij} = E \delta_{ij} + {1\over E} \left[ (a_L)^\mu p_\mu
- (c_L)^{\mu\nu} p_\mu p_\nu \right]_{ij} \,,
\label{eq:heff}
\ee
where $p_\mu = (E, -E\hat p)$ is the neutrino four-momentum, $\hat p$ the
neutrino direction, and $i,j$ are flavor indices. The coefficients
$a_L$ have dimensions of energy and the $c_L$ are dimensionless. The
Kronecker delta term on the right-hand side of Eq.~(\ref{eq:heff}) may be
ignored since oscillations are insensitive to terms in $h_{eff}$
proportional to the identity. For antineutrinos, $a_L \to -a_L$.

Direction dependence of the neutrino evolution enters via the space
components of $a_L$ and $c_L$. The coefficients may be specified in a
celestial equatorial frame $(T,X,Y,Z)$, which has $Z$
axis along the Earth's rotation axis and $X$ axis towards the vernal
equinox. The two-parameter bicycle model~\cite{K1}
can be defined as follows: $c_L$ is isotropic, with only one nonzero
element $(c_L)^{TT}_{ee} \equiv 2c$, and $(a_L)^\mu_{e\mu} =
(a_L)^\mu_{e\tau} = (0, a\hat Z/\sqrt2)$ are the only nonzero $a_L$.
We have generalized the model by letting $(a_L)^\mu_{e\mu} =
(a_L)^\mu_{e\tau} = (0, a\hat n/\sqrt2)$, where $\hat n$ is the 
preferred direction for the $a_L$ interaction. This increases the number of
parameters in the model to four, which is equal to the number required
in the usual massive neutrino description of oscillations (two mass-squared
differences and two mixing angles)~\cite{review}. We also consider
a five-parameter model which has a linear combination of direction-dependent
and direction-independent $a_L$.

In this letter we examine the phenomenology of this direction
dependence in the generalized bicycle model with massless
neutrinos. We find that the pure direction-dependent bicycle
model is ruled out by solar neutrino data alone, while a combination
of solar and long-baseline neutrino data excludes the
pure direction-independent case.  A mixture of direction-dependent and
direction-independent terms is excluded when KamLAND data are added.
In Sec.~2 we present the model and the neutrino oscillation
probabilities. In Sec.~3 we discuss the constraints from atmospheric
and long-baseline neutrino experiments.  In Sec.~4 we discuss the
constraints from solar neutrino experiments, and in Sec.~5 we discuss
the combined constraints, including KamLAND. In Sec.~6 we present our
conclusions.

\section{Neutrino oscillations in the generalized bicycle model}

Neutrino oscillations occur due to eigen energy differences in $h_{eff}$ and
the fact that the neutrino flavor eigenstates are not eigenstates of
$h_{eff}$. For massless neutrinos $p_\mu = (E, -E\hat p)$, where $\hat p$
is the direction of neutrino propagation. Then for the generalized
bicycle model
\be
h_{eff} = \pmatrix{
-2cE & {1\over\sqrt2}a\cos\Theta & {1\over\sqrt2}a\cos\Theta \cr
{1\over\sqrt2}a\cos\Theta & 0 & 0 \cr
{1\over\sqrt2}a\cos\Theta & 0 & 0} \,,
\ee
where
\be
\cos\Theta = \hat p\cdot\hat n \,,
\ee
{\it i.e.}, $\Theta$ is the angle between the neutrino momentum and the
preferred direction. From the diagonalization of $h_{eff}$, there are
two independent eigenenergy differences $\Delta_{jk} = E_j - E_k$,
\be
\Delta_{21} =
{m_0^2\over E_0^2} \left(\sqrt{E^2 + E_0^2 \cos^2\Theta} + E \right) \,,\quad
\Delta_{32} =
{m_0^2\over E_0^2} \left(\sqrt{E^2 + E_0^2 \cos^2\Theta} - E \right) \,,
\label{eq:Delta}
\ee
where $m_0^2$ and $E_0$ are defined in terms of the
Lorentz-invariance violating parameters by
\be
E_0 \equiv {a\over c} \,,\qquad m_0^2 \equiv {a^2\over c} \,,
\label{eq:params}
\ee
and the energy-dependent mixing angle is
\be
\sin^2\theta =
{1\over2}\left[ 1 - {E\over\sqrt{E^2 + E_0^2\cos^2\Theta}} \right] \,.
\label{eq:sin2theta}
\ee
The off-diagonal oscillation probabilities are~\cite{K1}
\bea
P(\nu_e \leftrightarrow \nu_\mu) &=& P(\nu_e \leftrightarrow \nu_\tau) =
2 \sin^2\theta \cos^2\theta \sin^2(\Delta_{31}L/2) \,,
\\
P(\nu_\mu \leftrightarrow \nu_\tau) &=&
\sin^2\theta \sin^2(\Delta_{21}L/2)
- \sin^2\theta \cos^2\theta \sin^2(\Delta_{31}L/2)
+ \cos^2\theta \sin^2(\Delta_{32}L/2) \,,
\label{eq:Pmt}
\eea
where $\Delta_{31} = \Delta_{32} + \Delta_{21}$.

If $E_0^2 \ll E^2$, {\it i.e.}, $a^2 \ll (cE)^2$, for atmospheric and long-baseline
neutrinos, then $\sin^2\theta \ll 1$, $\cos^2\theta \simeq 1$ and the only
appreciable oscillation is
\be
P(\nu_\mu \leftrightarrow \nu_\tau) \simeq \sin^2(\Delta_{32}L/2) \,,
\ee
where
\be
\Delta_{32} \simeq {m_0^2 \over 2E} \cos^2\Theta \,.
\label{eq:prob}
\ee
Thus the oscillation amplitude is maximal, the effective mass-squared
difference is
\be
\delta m^2_{eff} = m_0^2 \cos^2\Theta \,,
\label{eq:dm2}
\ee
and the energy dependence in this limit is the same as for
conventional neutrino oscillations due to neutrino mass differences.
Since the measured values for $\delta m^2_{eff}$ agree for atmospheric
neutrinos {\it and} the K2K~\cite{K2K} and MINOS~\cite{MINOS}
long-baseline experiments, the effective $\cos^2\Theta$ must also have
similar values in all of these experiments.

\section{Atmospheric and long-baseline neutrinos}

\subsection{Directional dependence}

\begin{figure}[t]
\centering\leavevmode
\includegraphics[width=4in]{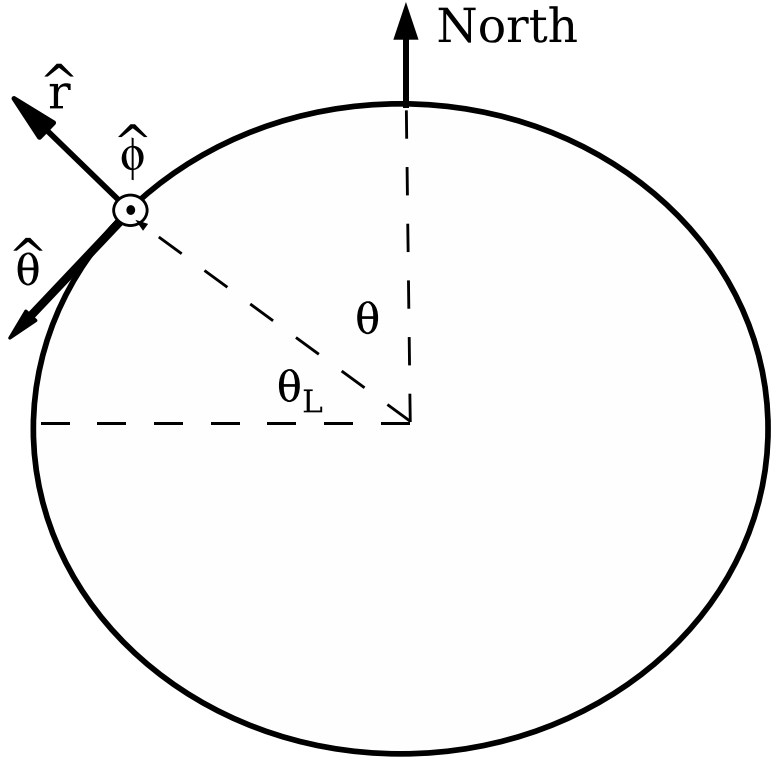}
\caption[]{Detector position in atmospheric and long-baseline
experiments. The angle $\theta_L$ is the latitude, while $\phi$ (not shown)
measures the time of the sidereal day. The unit vectors
$\hat r, \hat\theta, \hat\phi$ define the upward, southerly and easterly
directions, respectively, for a neutrino event in the detector.
\label{fig:coords}}
\end{figure}

With the Earth's rotation axis chosen as the $\hat Z$ direction and
the position of the detector given by $(\theta,\phi)$ in
a standard spherical polar coordinate system (see Fig.~\ref{fig:coords}),
the neutrino direction can be written as
\be
\hat p = -\cos\beta~\hat r + \sin\beta(-\sin\alpha~\hat\theta
+ \cos\alpha~\hat\phi) \,,
\ee
where ${\bf r}$ denotes the detector position, and the unit vectors
$\hat r$, $\hat\theta$ and $\hat\phi$ point in the upward, southerly and
easterly directions, respectively. The angle $\beta$ is the usual
zenith angle ($\beta = 0$ for a downward event) and $\alpha$ denotes the
compass direction of the neutrino velocity projected on the plane tangent
to the Earth's surface ($\alpha = 0$ for a neutrino going
in the eastward direction).  We take the preferred direction to be
\be
\hat n = \sin\xi\cos\chi~\hat X + \sin\xi\sin\chi~\hat Y + \cos\xi~\hat Z \,.
\ee
In our spherical polar coordinate system
\bea
\hat n &=& [\sin\xi\cos(\phi+\chi)\cos\theta_L +\cos\xi\sin\theta_L]~\hat r
+[\sin\xi\cos(\phi+\chi)\sin\theta_L-\cos\xi\cos\theta_L]~\hat\theta
\nonumber\\
&\phantom{=}& -\sin\xi\sin(\phi+\chi)~\hat\phi \,,
\eea
where the usual angle spherical polar $\theta$ has been replaced by the
latitude of the detector $\theta_L = {\pi\over2} -\theta$. (positive for
the northern hemisphere, negative for the southern hemisphere). The
azimuthal angle $\phi$ is chosen so that $\phi=0$ corresponds to the preferred
direction $\chi$, so that the angle $\chi$ may be dropped. The angular
dependence in the oscillation formulas is then
\bea
\cos\Theta &=& \cos\xi(\sin\beta\sin\alpha\cos\theta_L -\cos\beta\sin\theta_L)
\nonumber\\
&\phantom{=}& ~
-\sin\xi\cos\phi(\sin\beta\sin\alpha\sin\theta_L+\cos\beta\cos\theta_L)
-\sin\xi\sin\beta\cos\alpha\sin\phi \,.
\label{eq:cosT}
\eea
In Eq.~(\ref{eq:cosT}), $\xi$ gives the orientation of the preferred axis
with respect to the Earth's rotation axis, $\alpha$ (compass direction)
and $\beta$ (zenith angle) relate to the neutrino direction, and $\phi$
depends on the time of the sidereal day ($\phi = 0$ when the detector is
facing the preferred direction).

To help understand the complicated angular dependences in Eq.~(\ref{eq:cosT}),
we consider three special cases:
\bea
{\rm downward}(\beta = 0)&:& 
\cos\Theta = - (\cos\xi\sin\theta_L + \sin\xi\cos\theta_L\cos\phi) \,,
\label{eq:down}\\
{\rm upward}(\beta = \pi)&:& 
\cos\Theta = \cos\xi\sin\theta_L + \sin\xi\cos\theta_L\cos\phi \,,
\label{eq:up}\\
{\rm horizontal}(\beta = \pi/2)&:&
\cos\Theta = \cos\xi\cos\theta_L\sin\alpha
- \sin\xi(\sin\theta_L\cos\phi\sin\alpha+\sin\phi\cos\alpha)
\label{eq:side}\,.
\eea
Note that since only $\cos^2\Theta$ appears in the oscillation formulas,
the oscillation wavelengths for upward and downward events are the same.

\subsection{$\xi = 0$}

If the preferred direction is aligned with the Earth's rotation axis, then
$\xi = 0$ and
\be
\cos^2\Theta = (\sin\beta\sin\alpha\cos\theta_L - \cos\beta\sin\theta_L)^2 \,.
\ee
Note that in this case $\Theta$ does not depend on time of day (measured
by $\phi$). For accelerator experiments with relatively short baselines
compared to the Earth's radius (such as K2K and MINOS), the neutrino path
can be considered to be in the plane that is tangent to the Earth's surface,
so that Eq.~(\ref{eq:side}) applies and $\cos^2\Theta =
\sin^2\alpha\cos^2\theta_L$. 
Since the direction of the neutrino path in K2K is approximately
given by $\alpha \simeq 174^\circ$ (slightly north of west), and the latitude 
of the Super-K detector is $\theta_L \simeq 36.3^\circ$,
\be
m_0^2 = {\delta m^2_a \over \sin^2\alpha\cos^2\theta_L}
\simeq 0.4{\rm~eV}^2 \,.
\ee
For MINOS, $\alpha \simeq 124^\circ$ and $\theta_L \simeq
48^\circ$, so that
\be
m_0^2 \simeq 0.008{\rm~eV}^2 \,,
\ee
which is nearly two orders of magnitude smaller than the value required
to describe the K2K data. The reason K2K gives a much smaller value
for $\cos^2\Theta$ (and hence requires a much larger value for
$m^2_0$) is that the neutrino path is nearly perpendicular to the
Earth's rotation axis. Since the same $m_0^2$ applies to both, $\xi=0$
is excluded by a combination of the K2K and MINOS neutrino
experiments.

We note that for upward or downward atmospheric neutrino events,
$\cos^2\Theta = \sin^2\theta_L$, so $m_0^2 = \delta m^2_a/\sin^2\theta_L
\simeq 0.007$~eV$^2$, which is very close to the value extracted from
the MINOS data.

\subsection{$\xi \ne 0$}

If $\xi \ne 0$, then the preferred direction is not aligned with the
Earth's rotation axis. For upward or downward atmospheric events there
will be variation in $\cos^2\Theta$ (and hence in $\delta m^2_{eff}$)
over the sidereal period (see Eqs.~(\ref{eq:down}) and (\ref{eq:up})). At
the time of the sidereal day when $\phi = 0$ or $\pi$, there is always
an extremum in $\cos^2\Theta$. If $|\tan\xi| > |\tan\theta_L|$, then
there are two more extrema at $\cos\phi = -\tan\theta_L/\tan\xi$.
Thus there are two cases:

\begin{itemize}

\item For $|\tan\xi| < |\tan\theta_L|$, the only extrema of
$\cos^2\Theta$ occur at $\phi = 0$ and $\pi$. Specifically, if
$0 < \xi < \theta_L$, then there is a minimum at $\phi = 0$ and a
maximum at $\phi = \pi$, and
\be
\sin^2(\theta_L-\xi) \le \cos^2\Theta \le \sin^2(\theta_L+\xi) \,.
\label{eq:range1}
\ee
If $\pi - \theta_L < \xi < \pi$, then the positions of the maximum
and minimum reverse, and
\be
\sin^2(\xi+\theta_L) \le \cos^2\Theta \le \sin^2(\xi-\theta_L) \,.
\label{eq:range2}
\ee

\item For $|\tan\xi| > |\tan\theta_L|$ ({\it i.e.}, $\theta_L < \xi < \pi -
\theta_L$), $\cos^2\Theta = 0$ when $\cos\phi = -\tan\theta_L/\tan\xi$
(which occurs twice a day) and there are maxima at $\phi = 0$ and
$\pi$ with $\cos^2\Theta = \sin^2(\xi\pm\theta_L)$. Therefore
\be
0 \le \cos^2\Theta \le {\rm{max}}[\sin^2(\xi-\theta_L),\sin^2(\xi+\theta_L)] \,. 
 \label{eq:range3}
\ee

\end{itemize}

\begin{figure}[t]
\centering\leavevmode
\includegraphics[width=5in]{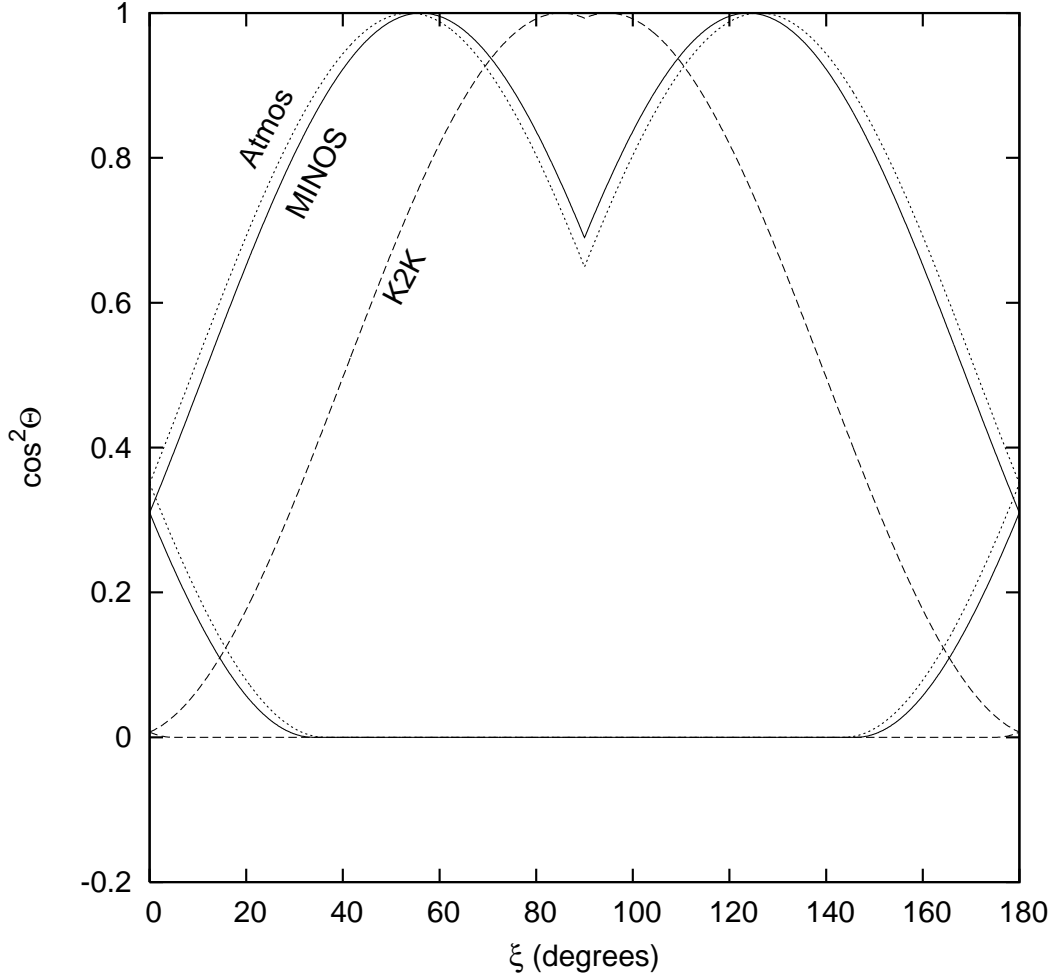}
\caption[]{
Maximum and minimum daily values for $\cos^2\Theta$ versus $\xi$ for
MINOS (solid curves), K2K (dashed) and Super-K up/down atmospheric (dotted)
data. In regions with two different local maxima, the larger one is shown.
\label{fig:dailyext}}
\end{figure}

The solid curves in Fig.~\ref{fig:dailyext} show the maximum and minimum
values of $\cos^2\Theta$ versus $\xi$ for upward and downward atmospheric
neutrinos. For $\theta_L < \xi < \pi-\theta_L$, there are always two times
during the sidereal day when $\cos^2\Theta = 0$, and hence
there are no oscillations for up/down events (since $\delta m^2_{eff} = m_0^2
\cos^2\Theta$). This effect might be evident in the Super-K data if it were
binned according to sidereal time.
For values of $\xi$ less than $\theta_L$ (or more than $\pi-\theta_L$),
$\cos^2\Theta$ is always finite, with the degree of modulation decreasing
as $\xi \to 0$ (or $\pi$).

There is a similar situation for horizontal events, except that the
critical angle that determines the number of extrema (and the values
for the extrema) is $\gamma = \sin^{-1}(\sin\alpha\cos\theta_L)$
instead of $\theta_L$.  For K2K, $\gamma \simeq 5^\circ$, and the
minimum $\cos^2\Theta$ is zero everywhere in the range $5^\circ < \xi<
175^\circ$. For $0 \le \xi \le 5^\circ$ and $175^\circ \le \xi \le
180^\circ$, the minimum $\cos^2\Theta$ is never larger than $\sin^25^\circ
\simeq 0.008$, so that there is always a time of day for K2K at which $\delta
m^2_{eff}$ is suppressed and there are effectively no
oscillations. The maximum and minimum $\cos^2\Theta$ for K2K are also
shown in Fig.~\ref{fig:dailyext}. For MINOS, $\theta_L = 47.8^\circ$
and the neutrino direction is approximately $\alpha = 124^\circ$; then
$\gamma = 34^\circ$ and the MINOS daily ranges for $\cos^2\Theta$ are
almost identical to those for Super-K up/down atmospheric events (see
Fig.~\ref{fig:dailyext}).

For $\xi < \pi/2$, the maximum value for $\delta m^2_{eff}$ in K2K is
$m_0^2\sin^2(\xi+\gamma_{K2K})$, and for $\xi \le \gamma_{MINOS}$, the
minimum value for $\delta m^2_{eff}$ in MINOS is
$m_0^2\sin^2(\xi-\gamma_{MINOS})$. For $\xi \le 12^\circ$ there is no value of
$m_0^2$ that gives $\delta m^2_{eff}$ within both allowed experimental
ranges ($1.9\times10^{-3}{\rm~eV}^2 \le \delta m^2 \le 3.5\times10^{-3}$~eV$^2$
for K2K and $2.3\times10^{-3}{\rm~eV}^2 \le \delta m^2 \le
3.4\times10^{-3}$~eV$^2$ for atmospheric neutrinos, at 90\%~C.~L.).\footnote{
This is approximately equal to the region where the $\cos^2\Theta$ values
do not overlap in Fig.~\ref{fig:dailyext}; the difference is due to the
slightly different ranges for $\delta m^2$ in the two experiments.} Therefore,
in an argument similar to the $\xi=0$ case, the predicted MINOS
and K2K $\delta m^2_{eff}$ disagree for
$\xi < 12^\circ$, in contradiction with data, so that these values are
excluded. For $12^\circ < \xi < 90^\circ$, there are always
two times during the sidereal day when $\cos^2\Theta = 0$ for K2K, and there
are no oscillations. For $12^\circ < \xi < 36^\circ$, atmospheric up/down
events should show a significant modulation of $\delta m^2_{eff}$, and for
$36^\circ < \xi < 90^\circ$ there are always two times during the sidereal
day when $\cos^2\Theta=0$ for atmospheric up/down events. Similar comments
can be made for the range $\pi/2 < \xi < \pi$.

The results for K2K, MINOS and up/down atmospheric neutrinos may be summarized
as follows:

\begin{itemize}

\item The range $0 < \xi < 12^\circ$ (and by similar arguments, $168^\circ
< \xi < 180^\circ$) is excluded by a comparison of the measured $\delta m^2$
values in MINOS and K2K data.

\item For $12^\circ < \xi < 168^\circ$, there are always two times
during the sidereal day when K2K should have no oscillations, {\it
i.e.}, no suppression of events relative to expectation. Up/down
atmospheric neutrinos always have a significant modulation of $\delta
m^2_{eff}$, and for $36^\circ < \xi < 144^\circ$ there are always two
times during the sidereal day when up/down atmospheric neutrinos
should also have no suppression.

\end{itemize}

For horizontal atmospheric neutrino events ($\beta = \pi/2$),
$\cos\Theta$ is given by Eq.~(\ref{eq:side}); the daily fluctuations then
depend on the compass direction of the event, $\alpha$. Super-K has
measured the compass dependence~\cite{compass} and found agreement with
an east-west asymmetry due to the Earth's magnetic field, plus oscillations.
Any additional compass dependence must not be too large to remain
consistent with the data. Table~\ref{tab:side}
shows $\cos^2\Theta$ for some typical values of $\alpha$ and $\phi$. The
direction dependence would not enhance or suppress the east-west difference,
but could enhance or suppress oscillations along the east/west direction
compared to north/south. Furthermore, enhancements could change to suppression
(and vice versa) during the sidereal period. A detailed analysis would be
needed to determine the compass-direction dependence for horizontal atmospheric
neutrino events.

\begin{table}[ht]
\caption[]{Values of $\cos^2\Theta$ for various values of $\alpha$ and $\phi$.
\label{tab:side}}
\vskip0.1in
\centering\leavevmode
\begin{tabular}{|l|cccc|}
\hline
$\alpha$ & $\phi = 0$ & $\phi = \pi/2$ & $\phi = \pi$ & $\phi = 3\pi/2$ \\
\hline
$0,\pi$ (E,W) & 0 & $\sin^2\xi$ & 0 & $\sin^2\xi$ \\
$\pi/2$, $3\pi\over2$ (N,S) & $\cos^2(\xi-\theta_L)$ & $\cos^2\xi\cos^2\theta_L$
& $\cos^2(\xi+\theta_L)$ & $\cos^2\xi\cos^2\theta_L$ \\
\hline
\end{tabular}
\end{table}

\section{Solar neutrinos}

\subsection{Directional dependence}

\begin{figure}[t]
\centering\leavevmode
\includegraphics[width=5in]{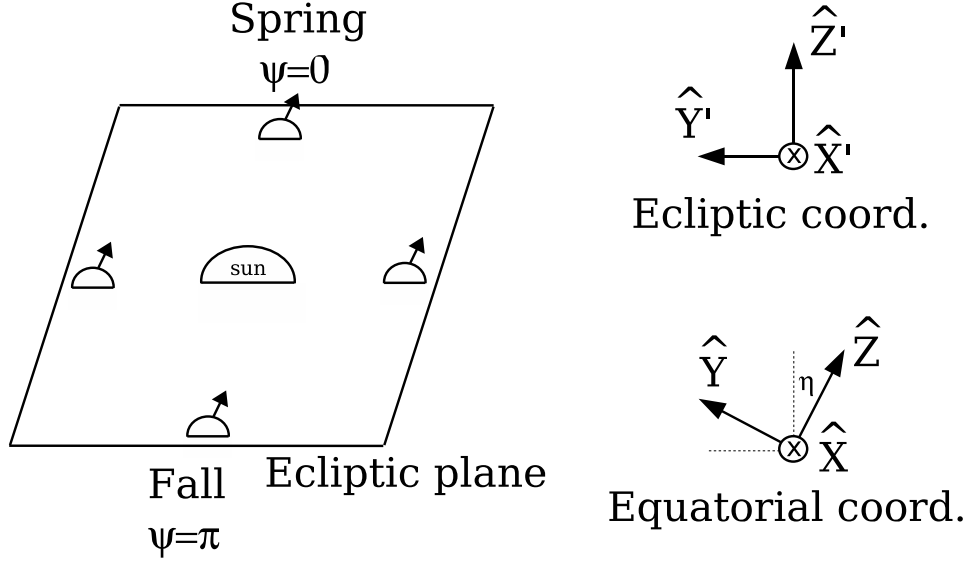}
\caption[]{
Position of the Earth in the ecliptic plane. The arrows represent the Earth's
rotation axis and $\psi = 0$ corresponds to the vernal equinox. The
orientation of the celestial equatorial coordinate system relative to the
ecliptic coodinate system is also shown.
\label{fig:ecliptic}}
\end{figure}

In a coordinate system $(X^\prime,Y^\prime,Z^\prime)$ where the $Z^\prime$
axis is perpendicular to the Earth's orbital plane (the ecliptic plane),
the direction of neutrino propagation may be written as (see
Fig.~\ref{fig:ecliptic})
\be
\hat p = \cos\psi~\hat X^\prime + \sin\psi~\hat Y^\prime \,,
\ee
where $\psi$ gives the position of the Earth in its orbit ($\psi=0$ at
the vernal equinox, $\psi=\pi/2$ at the summer solstice, etc.).
The equatorial coordinates are related to the ecliptic coordinates via
rotation by an angle $\eta \simeq 23^\circ$ about the $X^\prime$ axis,
where $\eta$ is the tilt of the Earth's rotation axis from the
perpendicular to the ecliptic (see Fig.~\ref{fig:ecliptic}). Then in the
celestial equatorial frame the direction of propagation for solar neutrinos is
\be
\hat p = \cos\psi~\hat X + \sin\psi\cos\eta~\hat Y - \sin\psi\sin\eta~\hat Z
\,,
\ee
and therefore
\be
\cos\Theta = \hat p \cdot \hat n
= \cos\psi\cos\chi\sin\xi + \sin\psi(\sin\chi\sin\xi\cos\eta-\cos\xi\sin\eta)
\,.
\label{eq:cosT2}
\ee
Note that $\cos\Theta$ for solar neutrinos is independent of detector latitude
($\theta_L$) and time of day ($\phi$).

\subsection{Oscillation probability}

In matter there is an additional term in the hamiltonian due to coherent
forward scattering of $\nu_e$'s with electrons in matter, so that
$-2cE$ in the upper left element of $h_{eff}$ is replaced by $-2cE
+ \sqrt2 G_F N_e$ and the mixing
angle in Eq.~(\ref{eq:sin2theta}) is then given by
\be
\sin^2\theta = {1\over2} \left[1 - {cE-G_FN_e/\sqrt2 \over
\sqrt{(cE-G_FN_e/\sqrt2)^2 + a^2\cos^2\Theta}} \right] \,,
\ee
where $N_e$ is the electron number density. For
adiabatic propagation in the sun the solar neutrino oscillation probability is
\be
P(\nu_e \to \nu_e) =
\cos^2\theta\cos^2\theta_0 + \sin^2\theta\sin^2\theta_0 \,,
\label{eq:P}
\ee
where $\theta_0$ is the mixing angle at the creation point in the sun (with
electron number density $N_e^0 \simeq 90 N_A$/cm$^3$) and $\theta$ is the
mixing angle in vacuum. $P \to {1\over 2}$ at 
low energies.\footnote{The actual 
solar neutrino oscillation probability at low energies
is closer to 0.7~\cite{BMW}. However, if an additional $a^{TT}_{ee}$ term
is included in $h_{eff}$, then the low energy probability can be fit to the
higher value, at the expense of adding a fifth parameter to the model.}
There is a minimum in $P$ at
\be
E_{min} = {G_F N_e^0\over2\sqrt2~c} \,,
\label{eq:Emin}
\ee
with minimum value
\be
P_{min}(\nu_e \to \nu_e) =  {4a^2\cos^2\Theta\over 8a^2\cos^2\Theta
+ (G_FN_e^0)^2} < {1\over2} \,,
\label{eq:Pmin}
\ee
where $\cos\Theta$ is given by Eq.~(\ref{eq:cosT2}). At $E = 2 E_{min}$
there is a resonance and the probablity is ${1\over2}$, and for
$E > 2 E_{min}$ the probability increases monotonically, with limiting
value unity as $E\to\infty$. The angle $\Theta$ depends on the time of
year; averaging over $\psi$ gives
\be
\langle P_{min}(\nu_e \to \nu_e) \rangle = {1\over2} \left[ 1 -
{G_F N_e^0\over\sqrt{(G_FN_e^0)^2 + 8 a^2 D^2}} \right] \,,
\label{eq:Pminavg}
\ee
where
\be
D^2 \equiv \cos^2\chi\sin^2\xi + (\sin\chi\sin\xi\cos\eta-\cos\xi\sin\eta)^2
\,.
\ee
If the probability minimum lies in the middle of the $^8$B solar neutrino
region, then $\langle P_{min} \rangle$ in Eq.~(\ref{eq:Pminavg}) will give the
approximate survival probability of the $^8$B neutrinos.

\begin{figure}[t]
\centering\leavevmode
\includegraphics[width=5in]{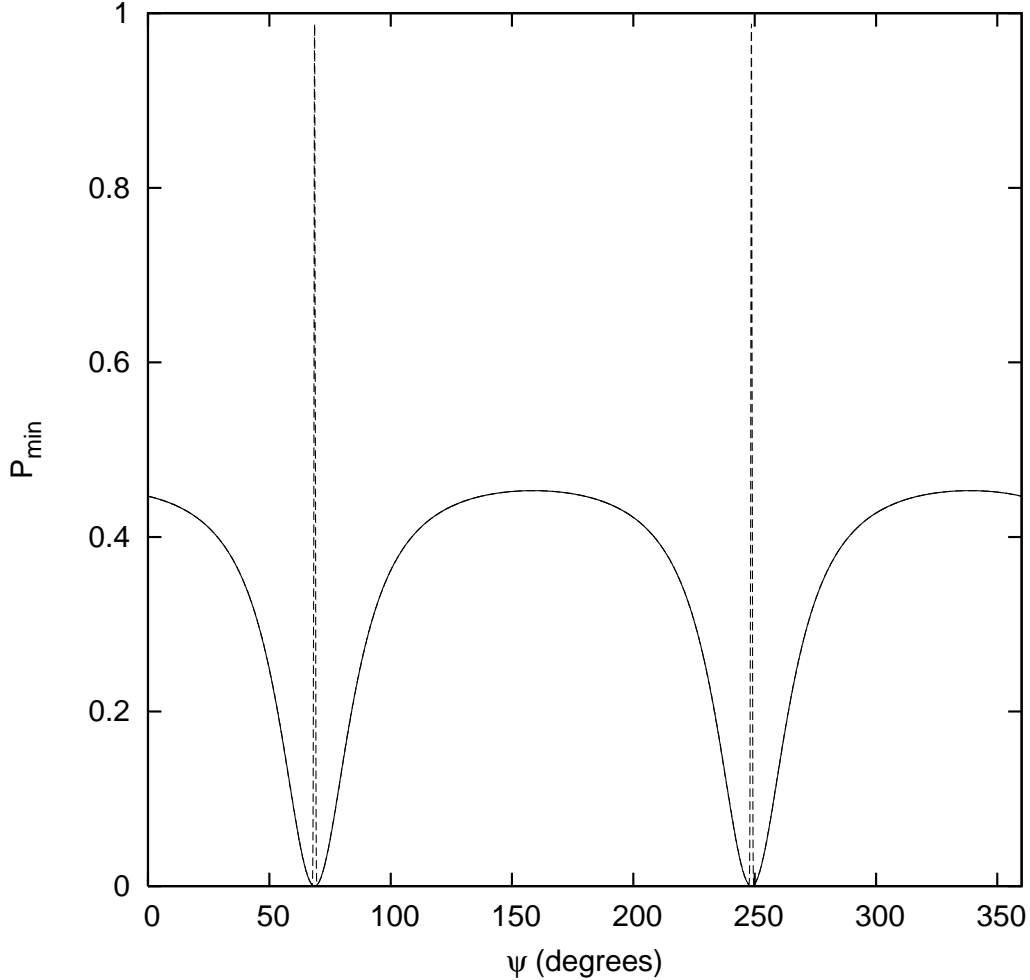}
\caption[]{
Representative solar neutrino survival probability at $E_{min}$ using
the adiabatic (Eq.~(\ref{eq:Pmin}), solid curve) and nonadiabatic
(Eq.~(\ref{eq:Pnonad3}), dashed) formulas, shown versus the time of year
(measured by $\psi$). The parameters for this example are $a =
7\times10^{-12}$~eV, $\xi = 45^\circ$ and $\chi = 0$, which give
$\langle P_{min} \rangle \simeq 0.34$. The two formulas differ
only close to the values of $\psi$ where $\cos\Theta = 0$.
\label{fig:Pnonad}}
\end{figure}

The formulas used above for the solar neutrino probability assumed adiabatic
propagation. It can be shown that the propagation is adiabatic except close
to the two times during the year where $\cos\Theta=0$:
\be
\psi = -\sin^{-1}\left(\cos\chi\sin\xi\over D \right) \quad{\rm and}\quad
\pi -\sin^{-1}\left(\cos\chi\sin\xi\over D \right)\,;
\label{eq:psi-nonad}
\ee
this was also pointed out in Ref.~\cite{K1} for the special case $\xi = 0$.
To include the effects of nonadiabatic propagation, Eq.~(\ref{eq:P}) must be
modified to
\be
P(\nu_e \to \nu_e) = {1\over2}\left[1 + (1-2P_x)
(\cos^2\theta\cos^2\theta_0 + \sin^2\theta\sin^2\theta_0) \right] \,,
\label{eq:Pnonad}
\ee
where $P_x$ is the level-crossing transition probability,
\be
P_x = e^{-\pi\gamma_r/2} \,,
\label{eq:Px}
\ee
and $\gamma_r$ is the adiabaticity of the transition at the level-crossing
resonance. For our Hamiltonian
\be
\gamma_r = {2 \sqrt2 a^2 \cos^2\Theta \over G_F |dN_e/dL|_r} \,,
\label{eq:gammar}
\ee
where $|dN_e/dL|_r$ is the rate of change of $N_e$ at the resonance.
At $E_{min}$ the probability becomes
\be
P_{min}(\nu_e \to \nu_e) =  {4a^2\cos^2\Theta\over 8a^2\cos^2\Theta
+ (G_FN_e^0)^2} + {(G_FN_e^0)^2\over(G_FN_e^0)^2+8a^2\cos^2\Theta} P_x \,,
\label{eq:Pnonad2}
\ee
where the first term on the right-hand side is the adiabatic contribution
and the second term the nonadiabatic correction. Propagation is
nonadiabatic when $\gamma_r$ is small, which occurs when
$a^2\cos^2\Theta$ is small. For the parameter ranges of interest we find
that $8a^2\cos^2\Theta \ll (G_FN_e^0)^2$ in the regions where $P_x$ is
nonnegligible, so that the probability reduces to
\be
P(\nu_e \to \nu_e) \simeq  {4a^2\cos^2\Theta\over 8a^2\cos^2\Theta
+ (G_FN_e^0)^2} + P_x \,.
\label{eq:Pnonad3}
\ee
From Eqs.~(\ref{eq:Px}), (\ref{eq:gammar}) and (\ref{eq:Pnonad3}) we see that
the survival probability goes to unity when $\cos\Theta = 0$.
The probability is shown in Fig.~\ref{fig:Pnonad} versus $\psi$ using both
the adiabatic and nonadiabatic formulas; they differ substantially only
near the values of $\psi$ given by Eq.~(\ref{eq:psi-nonad}).

\subsection{Constraints from solar data}

In order to fit the solar neutrino data, $\langle P_{min} \rangle$ must
match the measured probability for the $^8$B neutrinos, {\it i.e.},
$\langle P_{min}\rangle \simeq 0.34$ (we use the ratio of CC to NC rates
in SNO~\cite{SNO1} to avoid complications due to theoretical uncertainties
in the solar neutrino spectrum). Since there is no apparent energy dependence
in the $^8$B oscillation probability, the minimum must occur near the middle
of the $^8$B spectrum ($E_{min} \simeq 10$~MeV), so that probabilities at
either end of the spectrum are not much larger than in the middle. This
results in the two constraints (from Eqs.~(\ref{eq:Emin}) and
(\ref{eq:Pminavg}))
\bea
c &\simeq& {G_F N_e^0\over2\sqrt2~E_{min}} = 1.7\times10^{-19} \,,
\label{eq:c}\\
a D &\simeq&
\sqrt2 G_F N_e^0 {\sqrt{\langle P_{min}\rangle(1-\langle P_{min} \rangle)}
\over(1-2\langle P_{min}\rangle)}
= 5.0\times10^{-12}{\rm~eV} \,,
\label{eq:aD}
\eea
where Eq.~(\ref{eq:aD}) uses the adiabatic expression for $\langle P_{min}
\rangle$. Since Eq.~(\ref{eq:c}) depends only on the initial density for
$^8$B neutrinos and the central energy of the SNO spectrum, we will use
this result for $c$ throughout the rest of this paper.

We note that although the value of $a$ required to fit $\langle
P_{min} \rangle$ depends on the value of $D$ (which in turn depends on
the preferred-direction parameters $\xi$ and $\chi$), the product $aD$
is fixed by Eq.~(\ref{eq:aD}), and the oscillation probability versus
time will always be identical to that shown in Fig.~\ref{fig:Pnonad},
except for a possible shift in phase and the corrections for the
two nonadiabatic spikes. This can be understood by rewriting the adiabatic
probability in Eq.~(\ref{eq:Pmin}) as
\be
P_{min} = {4 a^2 D^2 \sin^2(\psi+\delta) \over 8 a^2 D^2 \sin^2(\psi+\delta) +
(G_FN_e^0)^2} \,,
\ee
where
\be
\tan\delta \equiv {\sin\xi\cos\chi \over \sin\chi\sin\xi\cos\eta -
\cos\xi\sin\eta} \,.
\label{eq:delta}
\ee
Thus the time variation of $P_{min}$ has the same shape and maximum and
minimum values when $aD$ is held fixed.

The measured solar neutrino survival probability does not exhibit much
variation throughout the year.  The SNO collaboration has tested their solar
neutrino data for periodicities~\cite{SNO2} and found a variation during the
year that is consistent with the $1/r^2$ dependence of the flux as the Earth's
distance from the sun varies.  The uncertainties in the rate are of
order 3-5\%, so there is little room for any additional annual
variation. The SNO periodicity data sample includes all of their solar
neutrino data in both the D$_2$O phase and salt phase, and combines
events from charge-current (CC), neutral-current (NC), electron
scattering (ES) and backgrounds (B). They measured the relative event
rate versus time of year, normalized to the mean rate, {\it i.e.},
\be
R = {N^0_{NC} + N^0_{CC}P + N^0_{ES}[P+r(1-P)] + N^0_B\over
N^0_{NC} + N^0_{CC}\langle P \rangle + N^0_{ES}[\langle P\rangle
+r(1-\langle P\rangle)] + N^0_B} \,,
\label{eq:R}
\ee
where $P$ is the oscillation probability, $N^0_i$ is the number of
events expected without oscillations, $r$ is the ratio of the NC to CC
cross sections and angle brackets indicate mean values. For $\langle
P_{min} \rangle = 0.34$, the bicycle model with directional dependence
predicts $R$ should vary between 0.42 and 1.19 throughout the
year. Since the SNO measurement of $R$ varies by at most 5\% at any
time during the year, the pure direction-dependent case clearly
cannot fit the SNO periodicity test while simultaneously reproducing
the correct average survival probabilty.

\begin{figure}[t]
\centering\leavevmode
\includegraphics[width=5in]{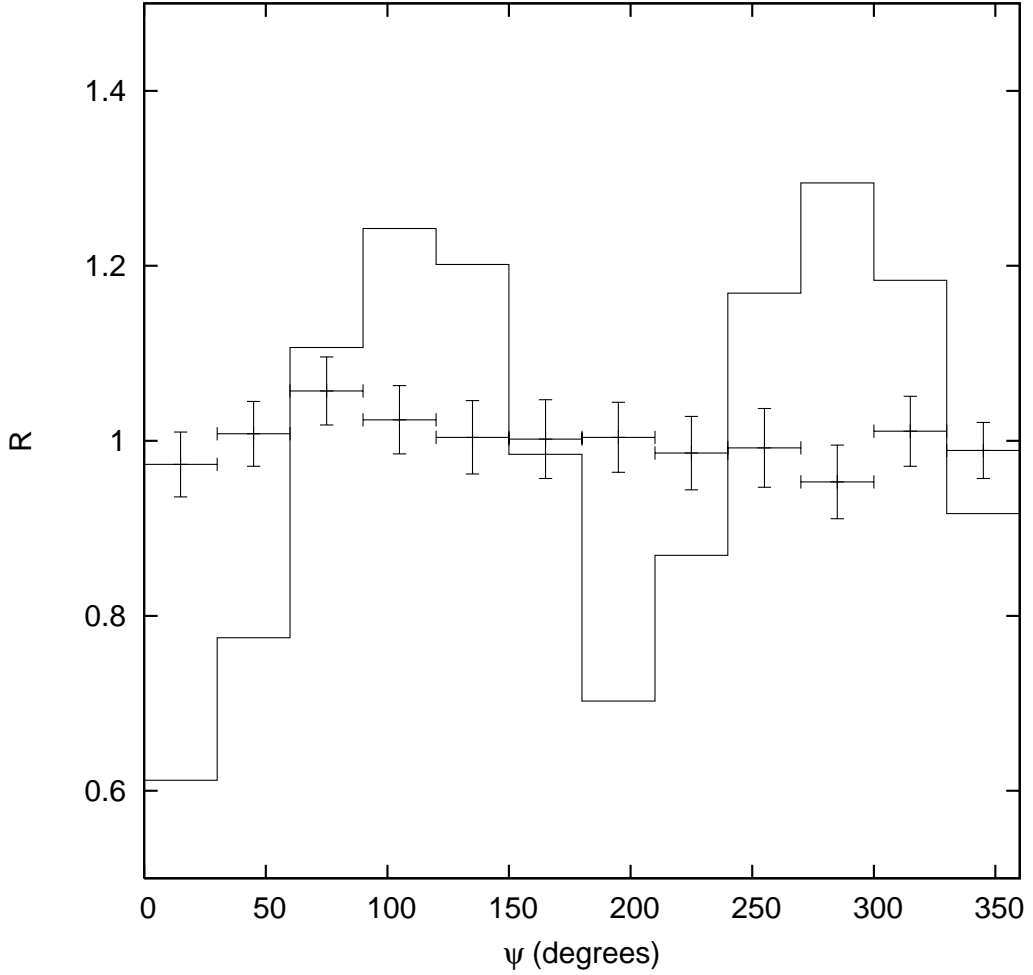}
\caption[]{
Best-fit prediction for $R$ in Eq.~(\ref{eq:R}) for the $^8$B
neutrinos (solid curve) and the SNO measured value for $R$
(data points) shown versus time of year (measured by $\psi$). Both the
SNO data and the model predictions are averaged over each bin, and
the SNO data has been corrected for the $1/r^2$ variation due to the
changing Earth-Sun distance. The prediction for $\langle P \rangle$ has
also been constrained to lie witin $1\sigma$ of the SNO central value.
The model parameters for the best fit are $a = 1.96\times10^{-12}$~eV$^2$,
$\xi = 43^\circ$ and $\chi=298^\circ$, with $\chi^2/DOF = 361/10$.
\label{fig:sno}}
\end{figure}

To verify this quantitatively we have searched the $a$, $\xi$ and
$\chi$ parameter space via Monte Carlo, using the twelve bins of the
SNO periodicity data and the SNO average probability ($0.34\pm0.03$,
from the CC to NC ratio).  We have used the appropriate weighting of
run times and D$_2$O/salt phase for each bin, and used
Eq.~(\ref{eq:Pnonad3}) for the oscillation probablity, which includes
the nonadiabatic part.  The nonadiabatic spikes appreciably affect
bin-averaged probabilities only in the bins where they occur, and then
by order 0.05 or less.  In Fig.~\ref{fig:sno} we show the SNO
periodicity data plus the best fit when varying over $a$, $\xi$ and
$\chi$ when $\langle P \rangle$ is constrained to lie within $1\sigma$
of the central value.  Allowing $\langle P \rangle$ to lie outside the
$1\sigma$ range can improve the $\chi^2$, but in all cases the
$\chi^2$ per degree of freedom (DOF) is such that the probablity that
the model describes the data is $2\times10^{-8}$ or less. The best fit
has very little annual variation, but $\langle P \rangle \simeq
0.21$. Therefore we conclude that the generalized direction-dependent
bicycle model is strongly ruled out solely by the solar neutrino
data.

\section{Combined constraints}

\subsection{Adding a direction independent term}

Since the pure direction-dependent case is ruled out, we now
generalize the model to include both direction-independent as well as
direction-dependent terms in the off-diagonal elements of
$h_{eff}$. This increases the number of parameters in the model to
five. If we define $(a_L)^\mu_{e\mu} = (a_L)^\mu_{e\tau} =
(a\cos\rho, a\sin\rho~\hat n/\sqrt2)$, where $\hat n$ is again the
preferred direction, then $\cos\Theta$ should be replaced by
$\cos\rho+\sin\rho\cos\Theta$ in our previous formulas. The parameter
$\rho$ determines the amount of direction dependence: $\rho = \pi/2$
or $3\pi/2$ corresponds to the pure direction-dependent case we
discussed before, while $\rho = 0$ or $\pi$ corresponds to no
direction dependence.

For a given preferred direction (fixed $\xi$ and $\chi$), the
parameters $c$ and $a$ are determined from the solar neutrino data
using Eqs.~(\ref{eq:Emin}) and (\ref{eq:Pmin}),
after the substitution $\cos\Theta \to \cos\rho+\sin\rho\cos\Theta$ is made.
Then using Eqs.~(\ref{eq:params}) and (\ref{eq:dm2}), $\delta m^2_{eff}$ for
long-baseline and atmospheric neutrinos may be written as
\be
\delta m^2_{eff} = {a^2\over c} (\cos\rho+\sin\rho\cos\Theta)^2 \,.
\ee
It is convenient to rewrite $\cos\Theta$ as
\be
\cos\Theta = D\sin(\psi+\delta) \,,
\label{eq:cosT3}
\ee
where $\delta$ is defined in Eq.~(\ref{eq:delta}). Integrating
$P_{min}$ in the modified Eq.~(\ref{eq:Pmin}) over $\psi$, leads to
\be
\langle P_{min} \rangle = {1\over2}\left[ 1 - {G_F N_e^0\over\sqrt2 S^2}
\sqrt{S^2 + 8a^2(D^2\sin^2\rho-\cos^2\rho) + (G_F N_e^0)^2} \right] \,,
\ee
for adiabatic neutrinos, where
\be
S^2 = \sqrt{64a^4(D^2\sin^2\rho-\cos^2\rho)^2 + 16a^2(G_F N_e^0)^2
(D^2\sin^2\rho + \cos^2\rho) + (G_F N_e^0)^4} \,.
\ee

\subsection{No direction dependence}

For the pure direction-independent case ($\rho = 0$ or $\pi$),
$\delta m^2_{eff} = m_0^2 = a^2/c$ for atmospheric and long-baseline
neutrinos and $P_{min}$
for $^8$B solar neutrinos is given simply by Eq.~(\ref{eq:Pmin}); for
$P_{min} = 0.34$,
\be
a = \sqrt{P_{min}\over2(1-2P_{min})}{G_F N_e^0 \over2}
\simeq 2.5\times10^{-12}{\rm~eV} \,,
\ee
and the prediction from the solar neutrino data is $\delta m^2_{eff} =
3.6\times10^{-5}$~eV$^2$ for atmospheric and long-baseline neutrinos, which is
clearly in contradiction with the data. Therefore, the pure
direction-independent case is ruled out by the combined data.

\subsection{Mixed case}

For a mixture of direction-dependent and direction-independent terms
in $h_{eff}$, a fit must be done to the solar data to determine an
allowed region in parameter space, and then the predictions for
$\delta m^2_{eff}$ in long-baseline experiments can be
compared to data. To fit the solar data we take the 12 bins from the SNO
periodicity data sample for the relative rate $R$ and add the additional
constraint that the average oscillation probability must be
$P = 0.34\pm0.03$, as described in the previous section. As before, we
fix the value of $c$ to that
given in Eq.~(\ref{eq:c}), and vary over the parameters $\xi$, $\chi$, $\rho$
and $a$ with a Monte Carlo. The 99\%~C.~L. allowed regions are determined
by restricting the $\chi^2/DOF$ to be less than 2.4 for nine DOF (there are
thirteen data points and four parameters).  The best fit to the SNO data 
has $a=3.0\times10^{-12}$~eV$^2$, $\xi=21^\circ$, $\chi = 94^\circ$ and
$\rho =114^\circ$, with $\chi^2/DOF = 4.84/9$.

\begin{figure}[t]
\centering\leavevmode
\includegraphics[width=5in]{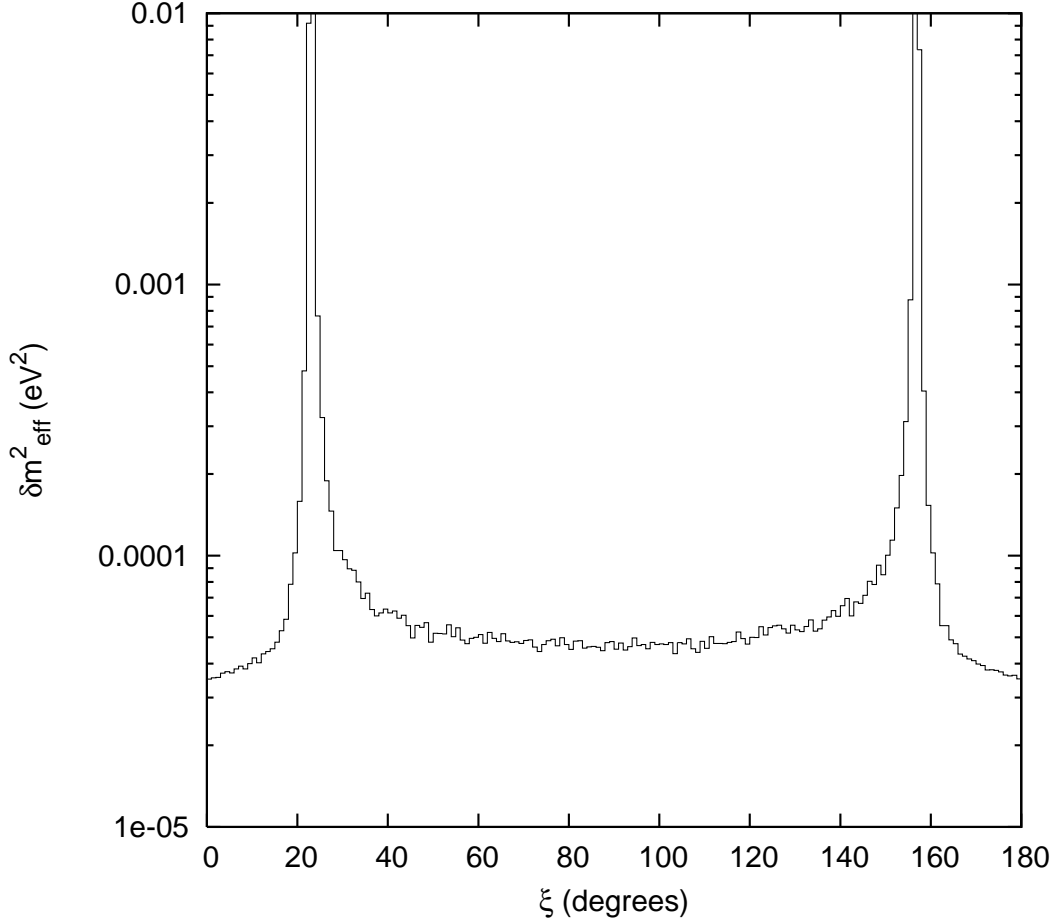}
\caption[]{
Maximum value of $\delta m^2$ in K2K allowed at 99\%~C.~L. by model parameters
consistent with the solar $R$ and $\langle P \rangle$ data, shown versus
the preferred direction $\xi$.
\label{fig:dm2}}
\end{figure}

Predictions for $\delta m^2_{eff}$ can then be made for K2K and MINOS.
Since $\delta m^2_{eff}$ depends on
$\cos\Theta$, it will vary during the sidereal day for $\xi \ne 0$,
with ranges depending on $\xi$ as shown in
Fig.~\ref{fig:dailyext}.  The strictest constraints come from K2K;
maximum possible values of $\delta m^2_{eff}$ in K2K are shown versus
$\xi$ in Fig.~\ref{fig:dm2}.

In most all cases the maximum possible $\delta m^2_{eff}$ can never be
in the experimentally measured range $1.9\times10^{-3} \le \delta m^2
\le 3.5\times10^{-3}$~eV$^2$.  Only a small region near $\xi \simeq
\eta = 23^\circ$ or $\pi-\eta \simeq 157^\circ$ can give a large
enough value of $\delta m^2_{eff}$. This allowed region is also
characterized by $\rho \simeq \pi/2$ or $3\pi/2$ and $\chi \simeq \pi/2$
or $3\pi/2$, such that $|D \sin\rho| \ll |\cos\rho|$, and values of
$a \ge 3\times10^{-11}$~eV.

As evident from Eq.~(\ref{eq:cosT3}), this results in $\cos\Theta
\simeq 0$ for solar neutrinos ({\it i.e.}, the preferred direction is
nearly perpendicular to the ecliptic plane), so that the directional
dependence for solar neutrinos is minimal, even though the
direction-dependent coefficient $\sin\rho$ is much larger than the
direction-independent coefficient $\cos\rho$. For atmospheric and
long-baseline neutrinos this fortuitous situation does not occur and
the direction-dependent piece is sizable, with daily variations of
$\cos^2\Theta$ given by Fig.~\ref{fig:dailyext}.  Therefore the case
with a mixture of direction-dependent and direction-independent terms
is severely constrained, and there is a strong variation of $\delta
m^2_{eff}$ for atmospheric and long-baseline neutrinos during the
sidereal day for the allowed solutions.

\subsection{KamLAND}

For reactor neutrinos with both direction-dependent and independent terms,
from Eqs.~(\ref{eq:Delta})-(\ref{eq:Pmt}) we have
\be
P(\bar\nu_e \to \bar\nu_e) =
1 - \sin^22\theta \sin^2\left({1\over2}\Delta_{31} L\right) \,,
\ee
where
\bea
\Delta_{31} &=& 2\sqrt{(cE)^2 + a^2(\cos\rho+\sin\rho\cos\Theta)^2} \,,
\\
\sin^22\theta &=& 1 - {(cE)^2 \over (cE)^2
+ a^2(\cos\rho+\sin\rho\cos\Theta)^2} \,,
\eea
and $\cos\Theta$ is given by Eq.~(\ref{eq:side}). For the values of the
parameters that fit solar data and give a large enough $\delta m^2_{eff}$
for long-baseline neutrinos, $(cE)^2 \ll a^2$ and
$\sin^22\theta \simeq 1$ at reactor neutrino energies, except possibly
for the brief time of day when $\cos\rho + \sin\rho\cos\Theta \to 0$.

As discussed in Sec.~3, given $\gamma =
\sin^{-1}(\sin\alpha\cos\theta_L)$, where $\alpha$ is the compass
direction of the incoming neutrino and $\theta_L$ the latitude of the
detector, the maximum and minimum values for $\cos^2\Theta$ during the
sidereal day are given by Eqs.~(\ref{eq:range1})-(\ref{eq:range3}) with
$\theta_L$ replaced by $\gamma$.  Then for the parameter ranges found
above, it is not hard to show that for all of the reactors
contributing to the KamLAND signal, the oscillation argument varies
over many cycles during the day, so that the oscillation probability
is close to 0.5, regardless of neutrino energy. Thus the bicycle model
gives a suppression in KamLAND that is nearly independent of energy,
contrary to the KamLAND data~\cite{kamland}, which excludes an
energy-independent suppression at 99.6\%~C.~L.

We have verified this result numerically using typical
solar/atmosperic/long-baseline solutions, averaging over the sidereal
day, and summing over individual reactor contributions -- the
suppression varies by at most 0.02 over the range $2.5 \le E \le
6$~MeV which supplies the bulk of the KamLAND data. The average
survival probability of the bicycle model solutions is at most about
0.55, well below the measured KamLAND value of $P = 0.658 \pm 0.044
\pm 0.047$. Therefore the bicycle model with a mixture of
direction-dependent and direction-independent terms is also excluded.

\section{Conclusions}

We have shown that the generalized five-parameter bicycle model with
Lorentz-invariance violation and no neutrino masses can be ruled out
by a combination of solar, long-baseline and reactor neutrino
data. The pure direction-dependent case is ruled out because it gives
a large annual variation in the oscillation probability for $^8$B
solar neutrinos, at odds with SNO periodicity data.  The pure
direction-independent case is ruled out because the values of the
parameters required to fit the SNO data predict a value of $\delta
m^2$ in long-baseline experiments that is too small by nearly two orders
of magnitude. Having a mixture of direction-dependent and
direction-independent terms in the off-diagonal elements of $h_{eff}$
is excluded when KamLAND is added to a combination of solar and
long-baseline data.

Although the five-parameter bicycle model cannot fit all of the data,
the full $h_{eff}$ with Lorentz-noninvariant oscillations of massless
neutrinos has 160 parameters~\cite{K1}, and a comprehensive comparison
with data is impractical. However, it is clear that any direction
dependence will encounter severe constraints, including variations
during the sidereal day which were not pursued in this paper.
Restricting $h_{eff}$ to only direction-independent terms reduces the
number of Lorentz-noninvariant parameters to 16~\cite{K1}.  Even then,
as our analysis of the direction-independent bicycle model suggests,
finding a set of parameters that would simultaneously fit solar,
atmospheric, long-baseline and reactor data will be difficult at best.

\section*{Acknowledgments}

We thank A. Kostelecky for useful discussions. We also thank the Aspen
Center for Physics for hospitality during the completion of this
work. This research was supported by the U.S. Department of
Energy under Grant Nos.~DE-FG02-95ER40896, DE-FG02-01ER41155, and
DE-FG02-04ER41308, by the NSF under CAREER Award No. PHY-0544278, and
by the Wisconsin Alumni Research Foundation.

\end{document}